# PREDICTIVE CYBER SECURITY ANALYTICS FRAMEWORK: A NON-HOMOGENOUS MARKOV MODEL FOR SECURITY QUANTIFICATION

Subil Abraham[1] and Suku Nair[2]

[1]IBM Global Solution Center, Coppell, Texas, USA
smabraha@us.ibm.com
[2]Southern Methodist University, Dallas, Texas, USA
nair@lyle.smu.edu

***ABSTRACT***

*Numerous security metrics have been proposed in the past for protecting computer networks. However we still lack effective techniques to accurately measure the predictive security risk of an enterprise taking into account the dynamic attributes associated with vulnerabilities that can change over time. In this paper we present a stochastic security framework for obtaining quantitative measures of security using attack graphs. Our model is novel as existing research in attack graph analysis do not consider the temporal aspects associated with the vulnerabilities, such as the availability of exploits and patches which can affect the overall network security based on how the vulnerabilities are interconnected and leveraged to compromise the system. Gaining a better understanding of the relationship between vulnerabilities and their lifecycle events can provide security practitioners a better understanding of their state of security. In order to have a more realistic representation of how the security state of the network would vary over time, a nonhomogeneous model is developed which incorporates a time dependent covariate, namely the vulnerability age. The daily transition-probability matrices are estimated using Frei's Vulnerability Lifecycle model. We also leverage the trusted CVSS metric domain to analyze how the total exploitability and impact measures evolve over a time period for a given network.*

## 1. INTRODUCTION

Defending a large scale enterprise from outside threats is a fairly complicated task. At the same time, Cybercriminals are increasingly using sophisticated social engineering techniques leading to disruptions in business operations, damaging the reputation as well as financial stability of these corporations. The recent cyber-attack incident at Target Corp illustrates how these security breaches can seriously affect profits and shareholder value. According to a report by Secunia[1], the number of reported security vulnerabilities in 2013 increased by 32% compared to 2012. However in spite of these increasing rate of attacks on corporate and government systems, corporations have fallen behind on ramping up their defenses due to limited budgets as well as weak security practices.

One of the main challenges currently faced in the field of security measurement is to develop a mechanism to aggregate the security of all the systems in a network in order assess the overall security of the network. For example INFOSEC [2] has identified security metrics as being one of the top 8 security research priorities. Similarly Cyber Security IT Advisory Committee [3] has also identified this area to be among the top 10 security research priorities.

In addition, traditional security efforts in corporations have focused on protecting key assets against known threats which have been disclosed publicly. But today, advanced attackers are developing exploits for vulnerabilities that have not yet been disclosed called “zero-day” exploits. So it is necessary for security teams to focus on activities that are beyond expected or pre-defined. By building appropriate stochastic models and understanding the relationship between vulnerabilities and their lifecycle events, it is possible to predict the future when it comes to cybercrime such as identifying vulnerability trends, anticipating security gaps in the network, optimizing resource allocation decisions and ensuring the protection of key corporate assets in the most efficient manner.

In this paper, we propose a stochastic model for security evaluation based on Attack Graphs, taking into account the temporal factors associated with vulnerabilities that can change over time. By providing a single platform and using a trusted open vulnerability scoring framework such as CVSS[4-6], it is possible to visualize the current as well as future security state of the network leading to actionable knowledge. Several well established approaches for Attack graph analysis [7-15] have been proposed using probabilistic analysis as well as graph theory to measure the security of a network. Our model is novel as existing research in attack graph analysis do not consider the temporal factors associated with the vulnerabilities, such as the availability of exploits and patches. In this paper, a nonhomogeneous model is developed which incorporates a time dependent covariate, namely the vulnerability age. The proposed model can help identify critical systems that need to be hardened based on the likelihood of being intruded by an attacker as well as risk to the corporation of being compromised.

The remainder of the paper is organized as follows. In Section 2, we discuss about previous research proposed for security metrics and quantification. In Section 3, we explore the Cyber-Security Analytics Framework and then realize a non-homogenous Markov Model for security evaluation that is capable of analyzing the evolving exploitability and impact measures of a given network. In Section 4, we present the results of our analysis with an example. Finally, we conclude the paper in Section 5.

## 2. Background and Related Work

Here we briefly discuss about Attack Graphs and then provide an overview of some of the most prominent works that have been proposed for quantifying security in a network.

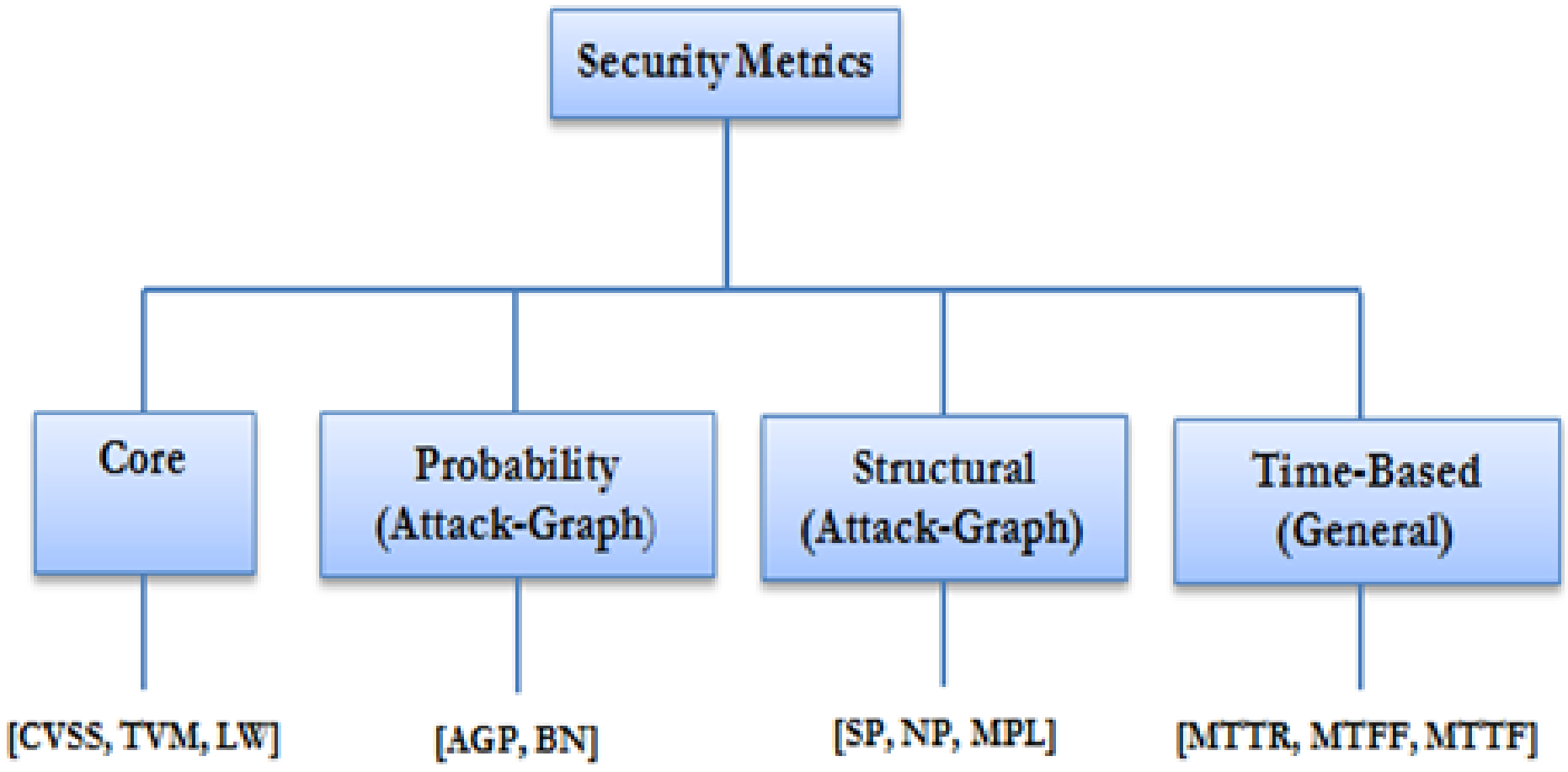

Figure 1. Security Metric Classification

## 2.1. Attack Graph

Computer attacks have been graphically modeled since the late 1980s by the US DoD as discussed in their paper [16]. Most of the attack modeling performed by analysts was constructed by hand and hence it was a tedious and error-prone process especially if the number of nodes were very large. In 1994 Dacier et al [17] published one of the earliest mathematical models for a security system based on privilege graphs. By the late 1990's a couple of other papers [18, 19] came out which enabled automatic generation of attack graphs using computer aided tools. In [18] the authors describes a method of modeling network risks based on an attack graph where each node in the graph represented an attack state and the edges represented a transition or a change of state caused by an action of the attacker. Since then researchers have proposed a variety of graph-based algorithms to generate attack graphs for security evaluation.

## 2.2. Classes of Security

There are different classes under which network security metrics fall under. These classes are depicted in Fig 1. Here are some examples of metrics that fall under each category.

### 2.2.1. Core Metrics

A few examples that fall under this category are Total Vulnerability Measure (TVM) [20] and Langweg Metric (LW) [21]. TVM is the aggregation of two other metrics called the Existing Vulnerabilities Measure (EVM) and the Aggregated Historical Vulnerability Measure (AHVM). CVSS [4-5] is an open standard for scoring IT security vulnerabilities. It was developed to provide organizations with a mechanism to measure vulnerabilities and prioritize their mitigation. For example the US Federal government uses the CVSS standard as the scoring engine for its National Vulnerability database (NVD) [6] which has a repository of over forty-five thousand known vulnerabilities and is updated on an ongoing basis

### 2.2.2. Structural Metrics

These metrics use the underlying structure of the Attack graph to aggregate the security properties of individual systems in order to quantify network security. The Shortest Path (SP) [18], [7] metric measures the shortest path for an attacker to reach an end goal. The Number of Paths (NP) [7] metric measures the total number of paths for an attacker to reach the final goal. The Mean of Path Lengths (MPL) metric [8] measures the arithmetic mean of the length of all paths to the final goal in an attack graph. The above structural metrics have shortcomings and in [9], Idika et al have proposed a suite of attack graph based security metrics to overcome some of these inherent weaknesses. In [22], Ghosh et al provides an analysis and comparison of all the existing structural metrics.

### 2.2.3. Probability-Based Metrics

These metrics associate probabilities with individual entities to quantify the aggregated security state of the network. A few examples that fall under this category are Attack Graph-based Probabilistic (AGP) and Bayesian network (BN) based metrics [23-25].

### 2.2.4. Time-Based Metrics

These metrics quantify how fast a network can be compromised or how quickly a network can take preemptive measures to respond to attacks. Common metric that fall in this category are Mean Time to Breach (MTTB), Mean Time to Recovery (MTTR) [26] and Mean Time to First Failure (MTFF) [27].

The drawback with all these classes of metrics is that they take a more static approach to security analysis and do not leverage the granularity provided by the CVSS metric framework in order to assess overall dynamic security situation and help locate critical nodes for optimization

### 2.3. Vulnerability Lifecycle Models

Presently, there is research [28-31] on analyzing the evolution of life cycle of different types of vulnerabilities. Frei et al. [31] in particular developed a distribution model to calculate the likelihood of an exploit or patch being available a certain number of days after its disclosure date. To the best of our knowledge, no previous work has been done to analyze overall security of a network, by considering the temporal relationships between all the vulnerabilities that are present in the network, which can be exploited by an attacker.

### 2.4. Cyber Situation Awareness

Tim Bass [32] first introduced the concept of cyberspace situation awareness and built a framework for it which laid the foundation for subsequent research in Network Security Situational Awareness [33]. The drawback with most of these NSSA models is that they don't adopt a consistent integrated framework for describing the relationships between the vulnerabilities in the network nor do they use an open scoring framework such as CVSS for analyzing the dynamic attributes of a vulnerability using stochastic modeling techniques.

## 3. Cyber-Security Analytics Framework

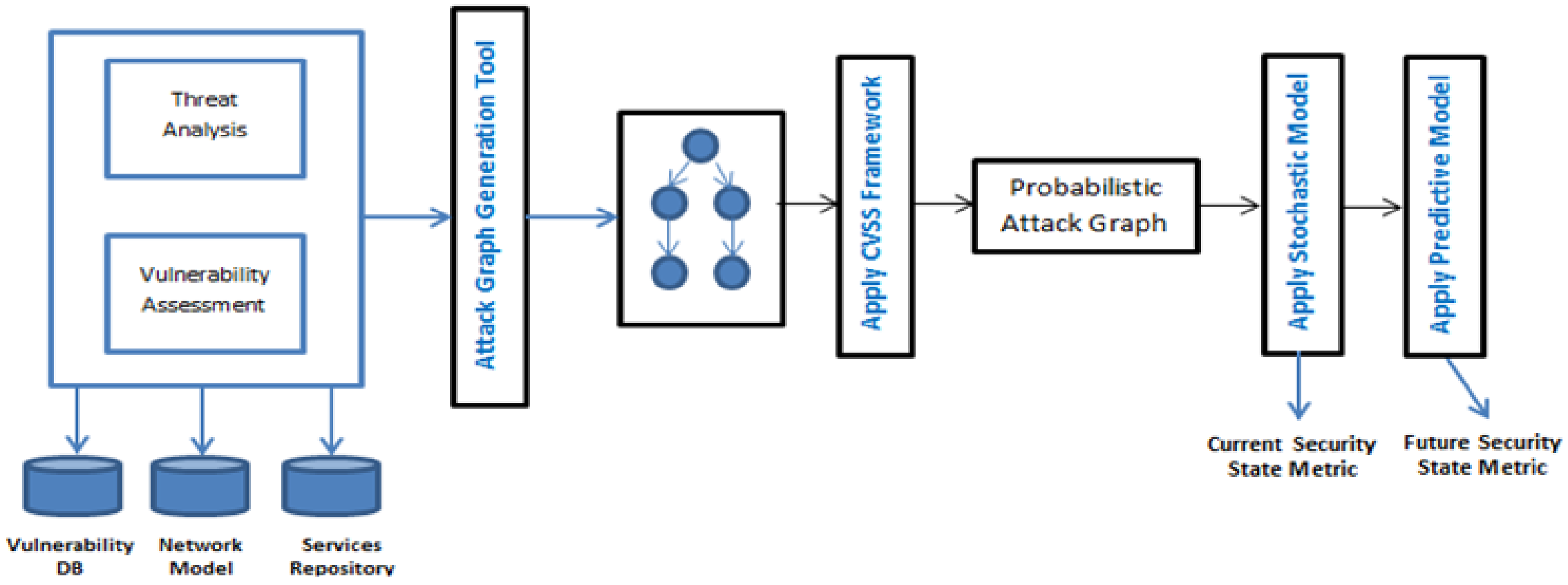

Figure 2. Cyber Security Analytics Framework

In this section, we explore the concept of modeling the Attack graph as a stochastic process. In [40, 41], we established the cyber-security analytics framework (Figure 2) where we have captured all the processes involved in building our security metric framework. In this paper we will extend the model by taking into account the temporal aspects associated with the individual vulnerabilities. By capturing their interrelationship using Attack Graphs, we can predict how the total security of the network changes over time. The fundamental assumption we make in our model is that the time-parameter plays an important role in capturing the progression of the attack process. Markov model is one such modeling technique that has been widely used in a variety of areas such as system performance analysis and dependability analysis [11, 27, 42-43]. While formulating the stochastic model, we need to take into account the behavior of the attacker. In this paper, we assume that the attacker will choose the vulnerability that maximizes his or her probability of succeeding in compromising the security goal

### 3.1. Architecture

Figure 3 shows a high level view of our proposed cyber security analytics architecture which comprises of 4 layers.

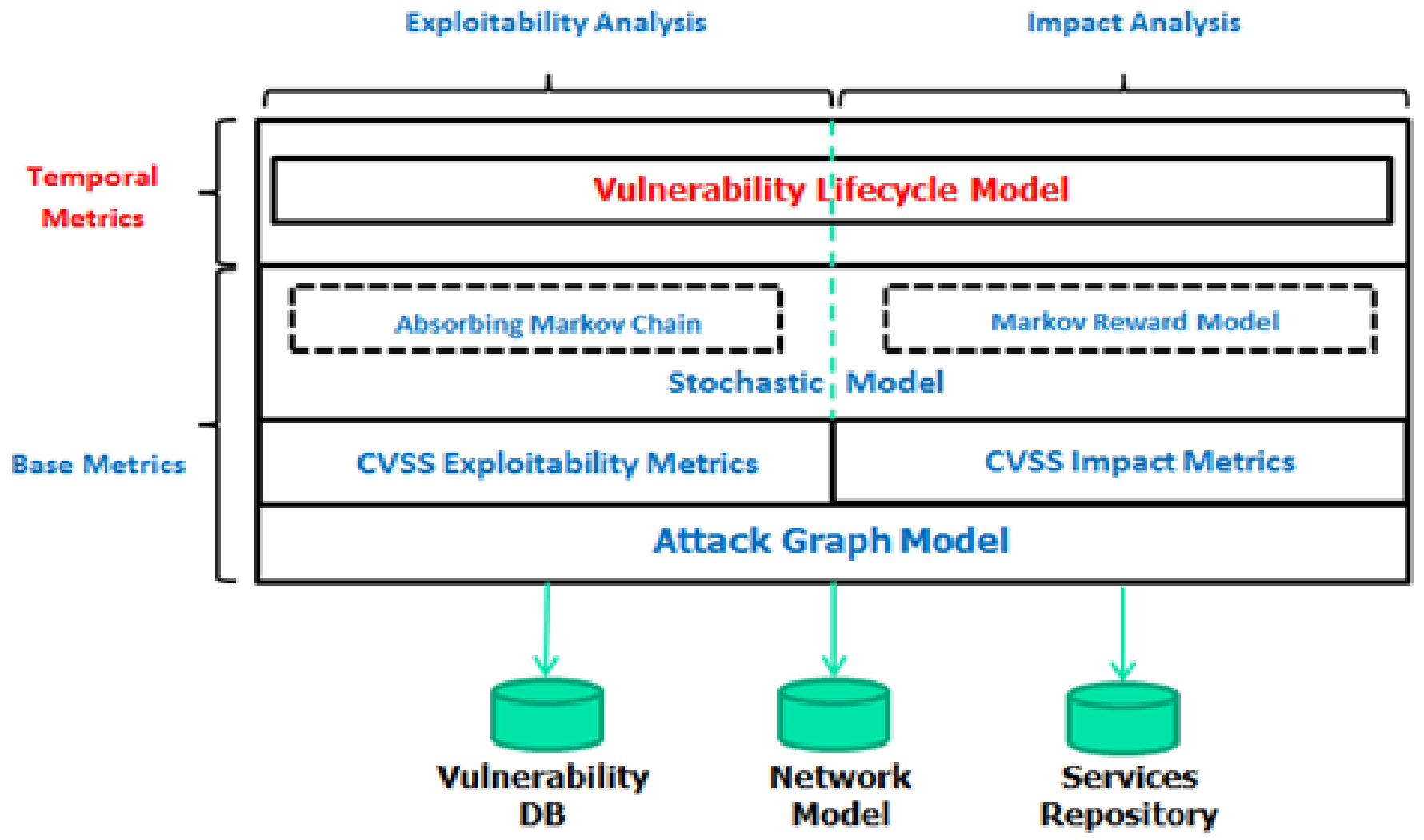

Figure 3. Cyber Security Analytics Architecture

The core component of our architecture is the Attack Graph Model (Layer 1) which is generated using a network model builder by taking as input network topology, services running on each host and a set of attack rules based on the vulnerabilities associated with the different services. The underlying metric domain is provided by the trusted CVSS framework (Layer 2) which quantifies the security attributes of individual vulnerabilities associated with the attack graph. We divide our security analysis by leveraging two CVSS metric domains. One captures the exploitability characteristics of the network and the other analyzes the impact a successful attack can have on a corporations key assets. We believe that both these types of analysis are necessary for a security practitioner to gain a better understanding of the overall security of the network. In layer 3, relevant stochastic processes are applied over the Attack Graph to describe the attacks by taking into account the relationships between the different vulnerabilities in a system. For example, in our approach, we utilize an Absorbing Markov chain for performing exploitability analysis and a Markov Reward Model for Impact analysis. In [40, 41] we discussed how we can model an attack-graph as a discrete time absorbing Markov chain where the absorbing state represents the security goal being compromised.

So far we have been focusing on the security properties that are intrinsic to a particular vulnerability and that doesn't change with time. These measures are calculated from the CVSS Base metric group which aggregates several security properties to formulate the base score for a particular vulnerability. In order to account for the dynamic/temporal security properties of the vulnerability, we apply a Vulnerability Lifecycle model (Layer 4) on the stochastic process to identify trends and understand how the security state of the network will evolve with time. Security teams can thus analyze how availability of exploits and patches can affect the overall network security based on how the vulnerabilities are interconnected and leveraged to compromise the system. We believe that such a framework also facilitates communication between security engineers and business stakeholders and aids in building an effective cyber-security analytics strategy.

### 3.2. Model Representation

In [40, 41] we have discussed how we can model an attack-graph as a discrete time absorbing Markov chain due to the following two properties.

1. An attack graph has at least one absorbing state or goal state.
2. In an attack graph it is possible to go from every state to an absorbing state.

We also presented a formula for calculating the transition probabilities of the Markov chain by normalizing the CVSS exploitability scores over all the transitions starting from the attacker's source state. In this paper, we will extend the model to analyze and measure two key aspects. First, we take into account both the exploitability as well impact properties associated with a security goal being compromised. By considering both these measures separately, we can derive a complementary suite of metrics to aid the security engineer in optimizing their decisions. Second we combine the temporal trends associated with the individual vulnerabilities into the model to reason about the future security state of the network. In our model we will calculate the daily transition-probability matrices using the well-established Frei's Vulnerability lifecycle model [31].

The security of the network is dependent on the exploitability level of the different vulnerabilities associated with the services running on the machines in the enterprise. In addition the security metrics will dynamically vary based on the temporal aspects of the vulnerabilities taken into consideration. As an example, consider CVE-2014-0416 which is an unspecified vulnerability in Oracle Java SE related to the Java Authentication and Authorization Service (JAAS) component. We define the base exploitability score $e(v)$ as the measure of complexity in exploiting the vulnerability $v$. The CVSS standard provides a framework for computing these scores using the *access vector* $(AV)$, *access complexity* $(AC)$ and *authentication* $(Au)$ as follows

$$e(v) = 20 \times AV \times AC \times Au \qquad (1)$$

The constant 20 represents the severity factor of the vulnerability. The access vector, authentication and access complexity for this vulnerability CVE-2014-0416 is 1.0, 0.704 and 0.71 respectively. Therefore the base exploitability score of CVE-2014-0416 is 10.0 which indicate that it has very high exploitability. As of this writing the state of exploitability for this vulnerability is documented as "Unproven" which indicate that no exploit code is available. Hence it has a temporal weight score of 0.85. Given the base exploitability score and the temporal weight, the effective temporal exploitability score is as follows

$$e(v_t) = temporal\,weight \times e(v) \qquad (2)$$

The temporal exploitability score for CVE-2014-0416 is 8.5. As the vulnerability ages and exploit code become readily availability to exploit the vulnerability, the value of the exploitability score will move closer towards its base value. As a comparison, consider CVE-2012-0551 which is another unspecified vulnerability in Oracle Java SE that has a base exploitability score of 8.6 which is lower than CVE-2014-0416. However the state of exploitability for this vulnerability is documented as "High" which indicates that exploit code is widely available. Hence it has a temporal weight score of 1. Therefore CVE-2012-0551 is considered more exploitable than CVE-2014-0416 even though it has a lower base metric score because we have factored in the lifecycle of the vulnerability.

The transition matrix for an absorbing Markov chain has the following Canonical form.

$$P = \begin{bmatrix} Q & R \\ 0 & I \end{bmatrix}$$

Here *P* is the transition matrix, *R* is the matrix of absorbing states, and *Q* is the matrix of transient states. The set of states, *S* in the model represent the different vulnerabilities associated with services running on the nodes that are part of the network. The transition probability matrix P of the Markov chain was estimated using the formula

$$p(i,j) = \frac{e(v_j)}{\sum_{k=1}^{n} e(v_k)} \qquad (3)$$

where $p_{ij}$ is the probability that an attacker currently in state i exploits a vulnerability $e(v_j)$ in state j and $e(v_j)$ is the exploitability score for vulnerability $v_j$ obtained from the CVSS Base Metric group. Further, each row of P is a probability vector, which requires that

$$\sum p(i,j) = 1 \; for \; all \; i \; \in S$$

In an absorbing Markov chain the probability that the chain will be absorbed is always 1. Hence

$$Q^n \to 0 \; as \; n \; \to \propto$$

where $Q$ is the matrix of transient states. Therefore for an absorbing Markov chain $P^{(m)}$, we can derive a matrix $N = (I - Q)^{-1} = I + Q + Q^2 + Q^3 + \cdots$ which is called the *fundamental matrix* for $P^{(m)}$. This matrix N provides considerable insight into the behavior of an attacker who is trying to penetrate the network. The elements of the fundamental matrix $n_{ij}$ describe the expected number of times the chain is in state *j*, given that the chain started in state *i*.

## 3.3. Non-homogenous model

The temporal weight score that we considered in the example above was a function of the time since the vulnerability was disclosed on a publicly trusted SIP (Security Information Providers). The temporal values recorded by CVSS are discrete and are not suitable for inclusion in a non-homogenous model. It would be more appropriate to use a distribution that would take as input the age of the vulnerability. Therefore we will use the result of Frei's model [31] to calculate the temporal weight score of the vulnerabilities that is part of the Attack graph model. The probability that an exploit is available for a given vulnerability is obtained using a Pareto distribution of the form:

$$F(t) = 1 - \left(\frac{k}{t}\right)^a \qquad (4)$$

a = 0.26, k = 0.00161

where t is the age of vulnerability and the parameter k is called the shape factor. The age t is calculated by taking the difference between the dates the CVSS scoring is conducted and when the vulnerability was first disclosed on an SIP. By combining this distribution in our model, we can get a more realistic estimate of the security of the network based on the age of the different vulnerabilities which are still unpatched in the enterprise.

In the non-homogenous Markov model presented here, we consider the following time dependent covariate which is the age of the vulnerability. In order to have a more realistic model, we update this covariate every day for each of the vulnerabilities present in the network and recalculate the discrete time transition probability matrix P. Given the exploitability scores for each of the vulnerabilities in the Attack Graph, we can estimate the transition probabilities of the Absorbing Markov chain by normalizing the exploitability scores over all the edges starting from the attacker's source state. Let $p_{ij}$ be the probability that an attacker currently in state i exploits a vulnerability in state j. We can then formally define the transition probability below where n is the number of outgoing edges from state i in the attack model and $e_j$ is the temporal exploitability score for the vulnerability in state j.

$$p(i,j) = \frac{e(v_t)_j}{\sum_{l=1}^{n} e(v_t)_l} \qquad (5)$$

$$where \; e(v_t) = \left(1 - \left(\frac{k}{t}\right)^a\right) \times e(v) \quad (6)$$

The matrix $P^{(m)}$ represents the transition probability matrix of the Absorbing Markov chain computed on day m where, $p\ (i,\ j) \geq 0$ for all $i,\ j\ \in S$. In an absorbing Markov chain the probability that the chain will be absorbed is always 1. Further, each row of $P^{(m)}$ is a probability vector, which requires that

$$\sum p(i,j) = 1\ for\ all\ i\ \in S$$

In equation (5), we use the result of Frei's Vulnerability Lifecycle model [31] to calculate the temporal weight score of the vulnerabilities that is part of the Attack graph model. Therefore by calculating the individual temporal scores of the vulnerabilities and by analyzing their causal relationships using an Attack Graph, we can integrate the effect of temporal score to derive the total dynamic security of network.

## 3.4. Exploitability Analysis

We present quantitative analysis using our cyber security analytics model. The focus of our analysis will on assessing the evolving security state of the network.

### 3.4.1 Expected Path length (EPL) metric

This metric measures the expected number of steps the attacker will have to take starting from the initial state to compromise the security goal. Using the Fundamental matrix N of the non-homogenous Markov model, we can compute the expected number of steps before the chain goes to the absorbed state. For example let $t_i$ be the expected number of steps before the chain is absorbed, given that the chain starts in state $s_i$, and let $t$ be the column vector whose $i^{th}$ entry is $t_i$. Then

$$t = Nc\ where\ for\ all\ j\ c_j\ is\ 1$$

This security metric is analyzing the expected number of steps or the resistance of the network.

### 3.4.2 Probabilistic Path (PP) metric

This metric measures the likelihood of an attacker to reach the absorbing states of the graph. For this we will calculate the following matrix B where $B = NR$ where N is the fundamental Matrix of the Markov chain and R is obtained from the Canonical form. The element $b_{ij}$ in the matrix measure the probability of reaching the security goal state j given that the attacker started in state i. The *Probabilistic Path (PP)* metric also aids the security engineer in making decisions on optimizing the network and we will label this as the *Probabilistic Path (PP)* metric.

## 3.4. Impact Analysis

The CVSS standard provides a framework for computing the impact associated with an individual vulnerability $v$ using *confidentiality impact* $(C)$, *integrity impact* $(I)$ and *availability impact* $(A)$ measures as follows

$$Impact(v) = 10.41 * (1 - (1 - C) * (1 - I) * (1 - A)$$

By associating the individual impact/reward scores with each state in our Markov chain, we can extend the underlying stochastic process as a discrete-time Markov reward model (MRM). Given our existing DTMC model, we can represent the Markov Reward process as $(\rho, S, P)$ where $(S, P)$ is a DMTC and $\rho$ is a reward function for each state. Since we are considering only constant rewards or impact scores, the reward function can be represented as a vector $r =\ [(\rho(s1), \ldots \rho(sn)]$. Hence the expected impact at time t is given as

$$\sum_{i=1}^{n} \rho(s_i).P\{X(t) = s_i\} = r.x(t) = r.P^t.x(0)$$

This value will be termed as the **Expected Impact Metric (EI)**. A non-homogenous MRM can be built by incorporating the temporal trend of individual vulnerabilities given their age. Hence by formulating daily transition probability matrices using Frei's model we can reason about how the expected cost of breaching a security goal can vary over time.

## 4. ILLUSTRATION

To illustrate the proposed approach in detail, a network similar to [10, 24-25, 44-45] has been considered (refer Figure 4).

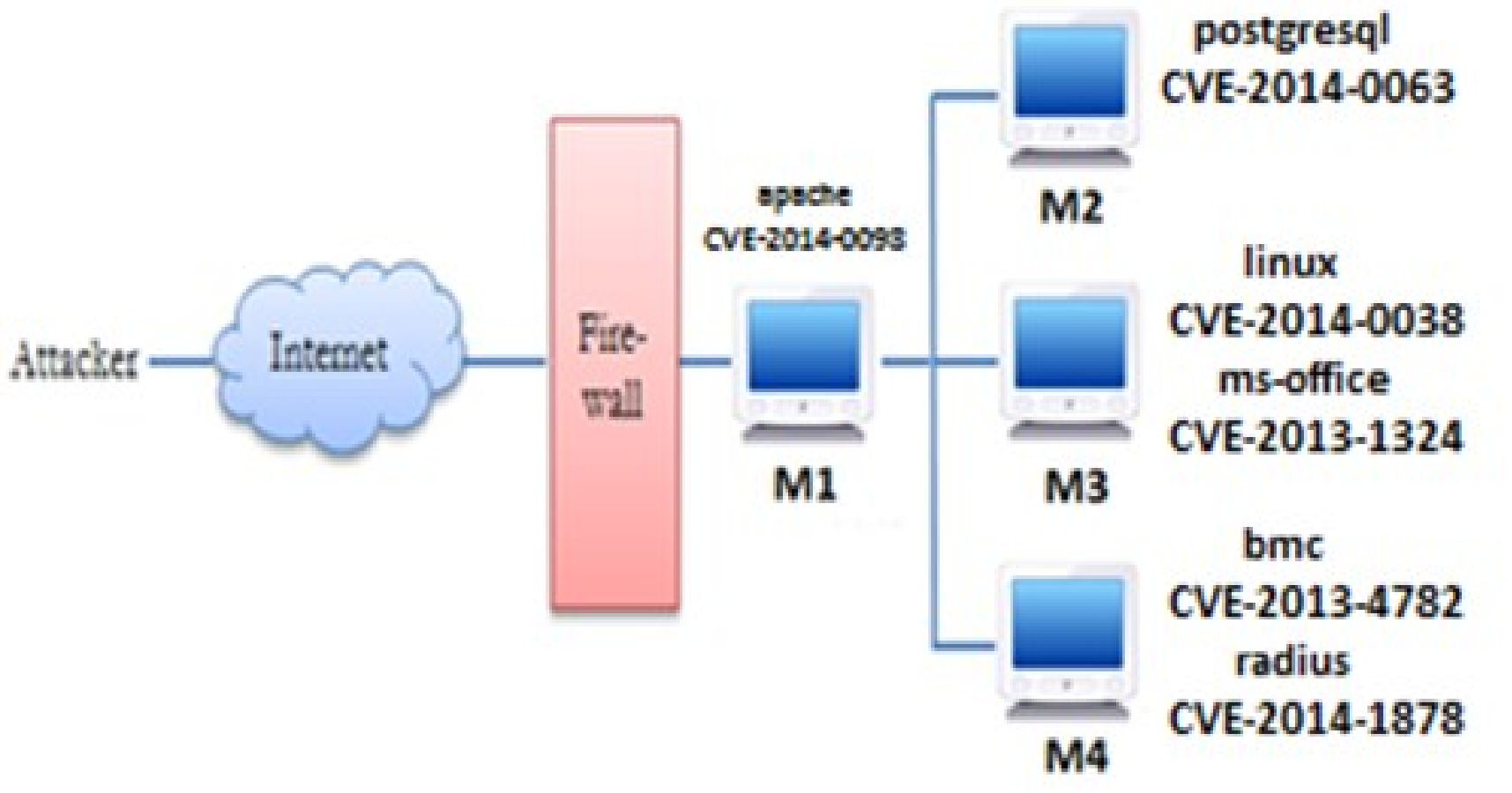

Figure 4. Network Topology

The network is comprised of 4 machines that are interconnected together and operating internally behind a firewall. The machine hosting the web-server M1 is running Apache Webserver. The aim of the attacker is to infiltrate the network and gain root access on M4. In order to accomplish this, the attacker needs to first start with exploiting the apache web-service since that is the only port (80) accessible from the firewall. Once this is exploited, the attacker will then need to slowly work his way through the network to achieve his goal.

### 4.1. Environment Information

Table 1 contains a list of all the vulnerabilities in the network that can be exploited by an attacker if certain conditions are met.

Table 1. Vulnerability Data.

| **Service Name** | **CVE-ID** | **Exploitability Subscore** | **Host** | **Disclosure Date** |
|---|---|---|---|---|
| apache | CVE-2014-0098 | 10.0 | M1 | 03/18/2014 |
| postgresql | CVE-2014-0063 | 7.9 | M2 | 02/17/2014 |
| linux | CVE-2014-0038 | 3.4 | M3 | 02/06/2014 |
| ms-office | CVE-2013-1324 | 8.6 | M3 | 11/13/2013 |
| bmc | CVE-2013-4782 | 10 | M4 | 07/08/2013 |
| radius | CVE-2014-1878 | 10 | M4 | 02/28/2014 |

Each of the six vulnerabilities is unique and publicly known and is denoted by a CVE (Common Vulnerability and Exposure) identifier. For example Apache web-server was found to have vulnerability CVE-2014-0098 on 03/18/2014 which allows remote attackers to cause segmentation faults. Similarly the postgresql service hosted by M2 had a vulnerability denoted by CVE-2014-0063 which allowed remote attackers to execute arbitrary code.

## 4.2. Attack Graph Generation

By combining the vulnerabilities present in the network configuration (Figure 4), we can build several scenarios whereby an attacker can reach a goal state. In this particular case, the attacker's goal state would be to obtain root access on Machine M4. Figure 5 depicts the different paths an attacker can take to reach the goal state. By combining these different paths we are able to obtain an Attack Graph. A couple of practical approaches have been proposed [44- 46] to automate generation of attack graphs without the intervention of a red team. In [47] the authors have compared and analyzed several open source AG tools like MulVal, TVA, Attack Graph Toolkit, NetSPA as well as commercial tools from aspects of scalability and degree of attack graph visualization. Table 2 provides an overview of the different available AG toolkits.

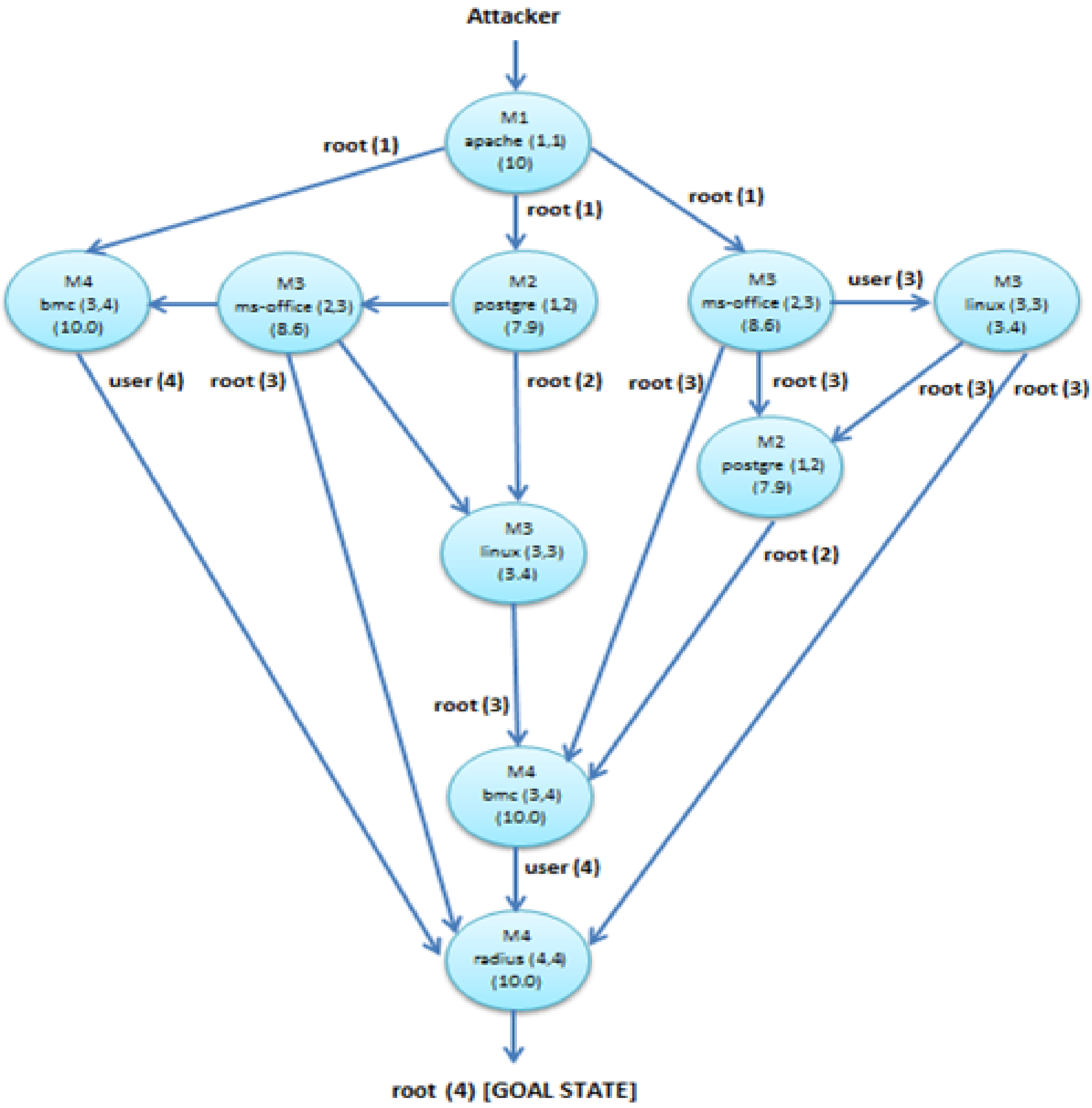

Figure 5. Network Topology

In our analysis we have used the MulVAL tool discussed in [45] to generate logical attack graph in polynomial time.

Table 2. Attack Graph Toolkits

| Toolkit Name | Complexity | Open Source | Developer |
|---|---|---|---|
| MulVAL | $O(n^2) \sim O(n^3)$ | yes | Kansas State University |
| TVA | $O(n^2)$ | no | George Mason University |
| Cauldron | $O(n^2)$ | no | Commercial |
| NetSPA | O(nlogn) | no | MIT |
| Firemon | O(nlogn) | no | Commercial |

## 4.3. Security Analysis

In the analysis, we first investigate how the distribution of our proposed attack graph metrics vary over a given time period. In the attack graph model, each node corresponds to a software related vulnerability that exists on a particular machine in the network. The transition probability for a particular edge in the attack graph is calculated by normalizing the CVSS Exploitability scores over all the edges from the attacker's source node. By formulating an Absorbing Markov chain over the Attack graph and applying the Vulnerability lifecycle model to the exploitability scores, we are able to project how these metrics will change in the immediate future.

Figure 6 (a, b, c) depicts the distribution of our proposed attack graph metrics (EPL, Probabilistic Path & Expected Impact) over a period of 150 days. The general trend for the Expected Path Length (EPL) metric (Fig 6a) is upward over the next 150 days which signifies that it will take fewer steps (less effort) for an attacker to compromise the security goal as the vulnerabilities in the network age. This visualization graph is very useful for security practitioners for optimizing patch management processes in the organization. By establishing thresholds for these metrics, the security teams can plan in advance as to when to patch a system versus working on an ad-hoc basis. For example, the organization may have a threshold score of 4.86 for the EPL metric. From the graph (Fig 6a), the team can reasonably conclude that the systems in their network are safe from exploits breaching the security goal for the next 50 days as the EPL score is above the threshold value. The thresholds values are typically set by the security team based on how fast they can respond to a breach once it is detected.

The Probability Path (PP) distribution (Fig 6b) which signifies the likelihood of an attacker compromising the security goal follows a different trend where it is seen reducing during the first few days and then picks up gradually with time. As vulnerabilities age, exploits become readily available for vulnerabilities which leads to an increase in their CVSS exploitability scores. As a result the transitions probabilities in the model will change likely causing other attack paths in the tree to become more favorable to the attacker. In the figure, we see that the probability of reaching the security goal tapers off after 50 days. The Expected Impact metric (Fig 6c) or cost of an attack to the business reduces with time and this is an indication that the attacker will likely choose a different path due to more exploits being available for other vulnerabilities that have a lower impact score. It is important to note that every organization has a network configuration that is very unique to their operations and therefore the distribution for our proposed attack graph metrics will be different for each of these configurations.

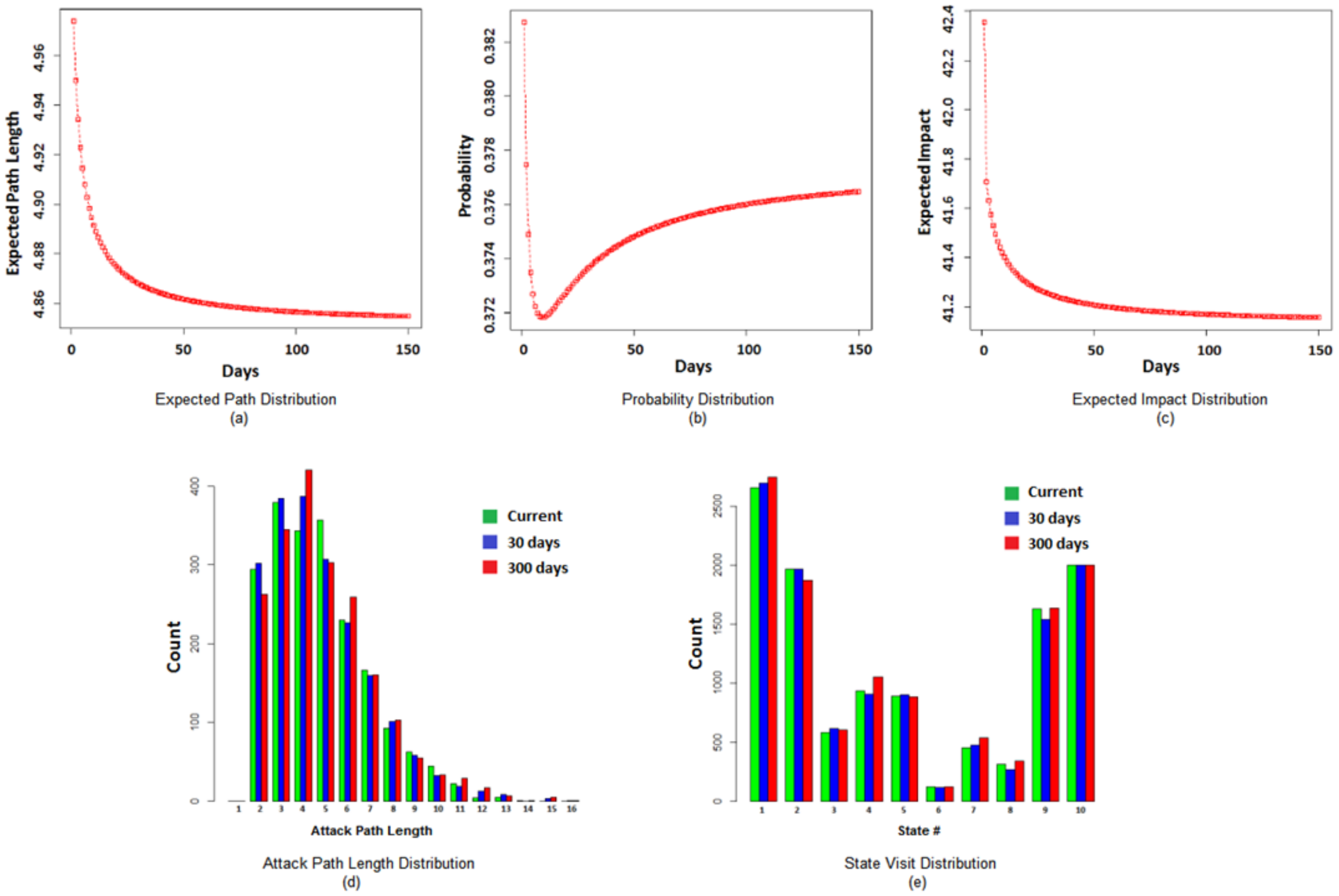

Figure 6. Network Topology

### 4.4. Simulation

Based on the Attack Graph generated for the network, a simulation of the Absorbing Markov chain is conducted. In our experiment we model an attacker and simulate over 2000 different instances of attacks over the Attack Graph based on the probability distribution of the nodes. We used the R statistic package [48] to generate the model and run the simulations.

The transition probabilities are formulated from the CVSS scoring framework as described in section 3. Each simulation run uses the transition probability row vector of the Absorbing Markov Chain to move from one state to another until it reaches the final absorbing state. Figure 6d depicts a multi-bar graph of the distribution of attack path lengths $X_1, X_2$ ..... $X_{2000}$ from 2000 simulated attack paths where the security goal was compromised. Each of the colors indicates the trend forecast over different periods of time. For example, in this distribution model, the length of the attack paths with the most frequency is 3, given the current age of all the vulnerabilities (shown in green). However as the vulnerabilities in the network age over a period of 300 days, the length of the paths with most frequency gets updated to 4 (shown in red). We also notice that the frequency of paths of length 3 and 5 decreases gradually over 300 days.

Similarly Figure 6e depicts a multi-bar graph of the distribution of the state visits for the Attack Graph for all the 2000 instances of attack paths that were simulated using the non-homogenous Markov chain model. This graph represents the number of times the attacker is likely to visit a particular state/node in the attack graph over those 2000 simulated runs. Based on the simulation results depicted in Fig 6e, if we were to exclude the start state (1) and the absorbing state (10), we can find that an attacker is most likely to visit state 2 and least likely to visit state 6. Hence from Table 1, we can conclude that the attacker is most likely to exploit the vulnerability of the bmc service running on M4 (State 2) and least likely to exploit the linux service on M3 (State 6). This information is valuable for a security engineer to prioritize which exploit needs to be patched and how it will affect the strength of the network against attacks. This insight is further enriched when we also consider the trends over a period of 300 days. For example, in Fig 6e there is an upward trend in the expected number of times the attacker visits state 4, while there is a downward trend for state 5 during the same time. Hence if the security engineer had to decide whether to patch node 4 or node 5 during this time period, it would make more sense to patch node 4 since it is most susceptible to an outside attack in the future.

One the major challenges when performing patch management is timing when to install patches and which patches have priority. By analyzing the trends over time of how the security state of a network changes, a security engineer can make a more informed and intelligent decision on optimizing the application of patches, thereby strengthening the current as well as the future security state of the enterprise.

## 5. CONCLUSIONS

In this paper, we presented a non-homogenous Markov model for quantitative assessment of security attributes using Attack graphs. Since existing metrics have potential short-comings for accurately quantifying the security of a system with respect to the age of the vulnerabilities, our framework aids the security engineer to make a more realistic and objective security evaluation of the network. What sets our model apart from the rest is the use of the trusted CVSS framework and the incorporation of a well-established Vulnerability lifecycle framework, to comprehend and analyze both the evolving exploitability and impact trends of a given network using Attack Graphs. We used a realistic network to analyze the merits of our model to capture security properties and optimize the application of patches.

## Authors

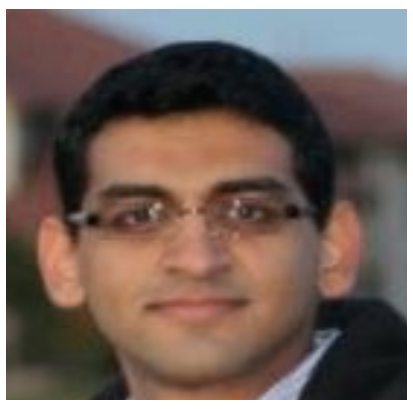

**Subil Abraham** received his B.S. degree in computer engineering from the University of Kerala, India. He obtained his M.S. in Computer Science in 2002 from Southern Methodist University, Dallas, TX. His research interests include vulnerability assessment, network security, and security metrics.

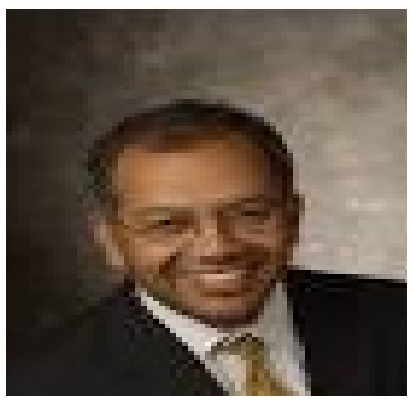

**Suku Nair** received his B.S. degree in Electronics and Communication Engineering from the University of Kerala. He received his M.S. and Ph.D. in Electrical and Computer Engineering from the University of Illinois at Urbana in 1988 and 1990, respectively. Currently, he is the Chair and Professor in the Computer Science and Engineering Department at the Southern Methodist University at Dallas where he held a J. Lindsay Embrey Trustee Professorship in Engineering. His research interests include Network Security, Software Defined Networks, and Fault-Tolerant Computing. He is the founding director of HACNet (High Assurance Computing and Networking) Labs. He is a member of the IEEE and Upsilon Pi Epsilon..